%% file: main.tex
\pdfoutput=1
\documentclass{aastex}
\usepackage{spr-astr-addons}
\usepackage{url}
\usepackage{color}
\usepackage[citecolor=blue, colorlinks=true, linkcolor=magenta]{hyperref}
\usepackage{amsmath, wasysym, amsfonts, amssymb, amsthm, bm}
\usepackage[section]{placeins}
\usepackage{graphicx, subfigure}
\usepackage{verbatim}
\usepackage[g]{esvect}
\usepackage{mathrsfs}
\usepackage{multirow, longtable, dcolumn, booktabs}
\usepackage{enumerate}
\usepackage{natbib}

\DeclareMathOperator{\sn}{sn}
\DeclareMathOperator{\cn}{cn}
\DeclareMathOperator{\dn}{dn}

\DeclareMathOperator{\ud}{d}

\DeclareMathOperator{\ii}{i}

\DeclareMathOperator{\sign}{sign}
\DeclareMathOperator{\km}{\,km}
\DeclareMathOperator{\degree}{^\circ}
\newtheorem{remark}{Remark}
\graphicspath{{./Figures/}}
\newcommand{\PreserveBackslash}[1]{\let\temp=\\#1\let\\=\temp}
\newcolumntype{C}[1]{>{\PreserveBackslash\centering}p{#1}}
\newcolumntype{R}[1]{>{\PreserveBackslash\raggedleft}p{#1}}
\newcolumntype{L}[1]{>{\PreserveBackslash\raggedright}p{#1}}

\begin{document}
\title{Analytic model for the long-term evolution of circular Earth satellite orbits including lunar node regression}
%
\shorttitle{Orbital plane evolution of circular Earth orbits}
\shortauthors{Zhu et al.}

\author{Ting-Lei Zhu\altaffilmark{1,2}} 
\and 
\author{Chang-Yin Zhao\altaffilmark{1,2}}
\and 
\author{Ming-Jiang Zhang\altaffilmark{1,2}}

\altaffiltext{1}{Purple Mountain Observatory, Chinese Academy Sciences, Nanjing, 210008}
\altaffiltext{2}{Key Laboratory of Space Object and Debris Observation, PMO, CAS, Nanjing, 210008}

%
\begin{abstract}
    This paper aims to obtain an analytic approximation to the evolution of circular orbits governed by the Earth's $J_2$ and the luni-solar gravitational perturbations.
    Assuming that the lunar orbital plane coincides with the ecliptic plane, \cite{Allan1964} derived an analytic solution to the orbital plane evolution of circular orbits.
    Using their result as an intermediate solution, we establish an approximate analytic model with lunar orbital inclination and its node regression be taken into account. Finally, an approximate analytic expression is derived, which is accurate compared to the numerical results except for the resonant cases when the period of the reference orbit approximately equals the integer multiples (especially 1 or 2 times) of lunar node regression period.
\end{abstract}
\keywords{Orbital plane evolution; Milankovitch element; Lunar node regression; Analytic approximation}

%
\input{Introduction}
\input{Model}
\input{Solution}

\input{Conclusion}
%

\acknowledgments
This research was supported by the Youth Program of the National Natural Science Foundation of China (Grant No. 11603078) and the Key Program of the National Natural Science Foundation of China (Grant No. 11533010). 
The authors would like to thank two anonymous reviewers for the valuable comments that help to substantially improve the manuscript.
The authors would also like to thank Dr. Fangzhou Jiang for his help in improving the language.
The first author would like to give special thanks to his bride, Q. Qian, for her endless support.

%
\bibliographystyle{apa}
\bibliography{2016_orbital_plane}
\end{document}

%% file: Introduction.tex
\section{Introduction}
The problem of the orbital plane evolution of a satellite around an oblate planet has been studied since two hundred years ago when Laplace investigated the motion of Iapetus, the third largest natural satellite of Saturn.
Sixty years ago, the launch of Sputnik-1 boosted similar research in the field of artificial satellite theory.
\citet{Allan1964} presented their analytic results on the long period motion of the plane of distant circular orbits due to the Earth's oblateness and the luni-solar gravitational perturbations. The dynamical system was doubly averaged over the mean motions of both the satellite and the disturbing bodies, and further simplified by the assumption that the moon is orbiting the Earth in the ecliptic plane.  
Then the classical Laplace plane was defined as the local equilibrium of the simplified dynamical system. 
In addition, according to the analytic results, the orbital plane of a circular orbit was precessing around the Laplace plane, with a period of several decades depending on the semi-major axis and the initial direction of the orbital pole.

A number of theories and applications were presented after the classical work of \citet{Allan1964}.
Satellite dynamics on the Laplace surface of general Sun-Planet-Satellite systems were discussed in detail by \citet{Tremaine2009}.
The Laplace plane at the Geosynchronous Earth Orbit (GEO) altitude (i.e. $a\approx{42164}{\km}$) is approximately $7.4\degree$ inclined to the equator, which implies that the inclination of conventional GEO satellites ($a\approx{42164}{\km}$, $e\approx 0$ and $i\approx{0}\degree$) 
varies in the range of $[{0}{\degree}, {15}{\degree}]$ with a period of about 53 years \citep{Valk2009,Ulivieri2013,Zhao2015}.
The orbital plane cannot be frozen in the classical Laplace plane, and the periodicity of orbital plane evolution cannot reveal real motion, when the analysis is extended to Medium Earth Orbit (MEO), due to the influence of Moon's real motion, especially the lunar orbital regression along the ecliptic plane \citep{Ulivieri2013,Circi2016}.

\citet{Allan1964} warned that circular motion may be unstable for some inclinations due to the Kozai-Lidov effect or luni-solar resonance in general \citep{Breiter2001}.
Recently, a series of papers showed that the eccentricity of the orbits for Global Navigation Satellite System (GNSS) may increase exponentially \citep{Chao2004,Bordovitsyna2012}.
Through theoretical calculations and numerical verifications for the Chinese BeiDou Inclined Geosynchronous Earth Orbit (IGSO), we found that the eccentricity of the near-circular IGSO satellite with an initial inclination not larger than $32\degree$ can always remain small in the long-term evolution, and the initial Right Ascension of the Ascending Node (RAAN) can affect the eccentricity too \citep{Zhao2015}.
\citet{Daquin2016a,Daquin2016} and \citet{Gkolias2016} improved the investigation of eccentricity excitation by introducing dynamical indicators (e.g. Fast Lyapunov Indicator, or FLI) to describe the chaoticity of the dynamic systems for the highly inclined orbits. 
Their results verified that the instability is highly correlated with the orbital plane and demonstrated the complexity of the problem. 

Therefore, an investigation of the orbital plane variation of circular orbits with the consideration of lunar node regression is crucial in understanding the orbital dynamics.
This paper aims to obtain an analytic approximation to the evolution of circular orbits governed by the Earth's oblateness and the luni-solar gravitational perturbations. The main body of this paper is organised as follows:

Section \ref{sec:model} briefly introduces the doubly averaged dynamic model in terms of vectors and tensors. 
Similar to \citet{Circi2016}, the force model consists of the Earth's $J_2$ perturbation and the luni-solar gravitational perturbations. 
The lunar node regression is also taken into consideration.
Then the force model is divided into an autonomous part and a non-autonomous part (seen as the perturbation), where the former can be solved analytically by the method presented in \citet{Allan1964}. 
Treating their solution as an intermediate orbit, we introduce a variable transformation and establish the equation of motion using the method of constant variation. 
Section \ref{sec:solution} presents the analytic solution of the doubly averaged model.
A general expression describing the long-period motion of the orbital plane is derived from the equation of motion, which is suitable for non-resonant circular orbits with small or medium inclination, except for those close to the classical Laplace plane.
The analytic approach for the quasi-frozen (i.e. near Laplace plane) orbits is accomplished based on the linearisation of the original equation.
Section \ref{sec:numerical} applies massive numerical simulations to verify the analytic model.

%% file: Model.tex
\section{Doubly averaged model}
\label{sec:model}
The long-term evolution of a near-circular orbit can be described by its orbital angular momentum, and the equation of motion takes the form of \citep{Allan1964,Circi2016}
\begin{equation}
    \dot{\bm R} = \bm R\times \bm{MR}
\label{eq:motion}
\end{equation}
where $\bm R$ is a unit vector along the orbital angular momentum, and $\bm M$ is a two dimensional symmetric tensor
\begin{equation}
    \bm M=\sum_{j=0}^{2}\omega_j\bm Z_j\bm Z_j^T
\end{equation}
with $\bm Z_0$, $\bm Z_1$ and $\bm Z_2$ unit vectors along the Earth pole, the ecliptic pole and the lunar orbital angular momentum, respectively.
$\bm Z_j\bm Z_j^T$ are symmetric dyads (refer to the appendices of \cite{Rosengren2014c}).
The coefficients $\omega_j$'s take the following form
\begin{equation}
\left\{
\begin{aligned}
    & \omega_0 = \frac{3}{2}\frac{nJ_2R_e^2}{a^2} \\
    & \omega_1 = \frac{3}{4}\frac{\mu_S}{na_S^3(1-e_S)^{3/2}} \\
    & \omega_2 = \frac{3}{4}\frac{\mu_M}{na_M^3(1-e_M)^{3/2}} 
\end{aligned}
\right.
\end{equation}
where $R_e$ is the Earth's equator radius; $\mu_S$ and $\mu_M$ are the gravitational constants of the sun and the moon, respectively; $a_S$, $e_S$, $a_M$ and $e_M$ are the Keplerian orbital elements of the sun and the moon in the Geocentric Ecliptic Reference System (GERS, $O\text{\,-\,}X_1Y_1Z_1$); $a$ is the orbital semi-major axis of the satellite and $n$ is the satellite's mean motion.
Obviously, $\omega_j$ are one variable functions of semi-major axis, $a$.
In addition, in the doubly averaged dynamical system, $a$ is a first integral (or constant of motion), then $\omega_j$ can be seen as constants.

Given some reference system, the vectors $\bm Z_j$ can be expressed by $3\times 1$ matrices and thus the symmetric tensor $\bm M$ can be expressed by $3\times 3$ matrix.
In GERS, 
\begin{equation}
\left\{
    \begin{aligned}
        & \bm Z_0 = (0, \sin\varepsilon,\cos\varepsilon)^T \\
        & \bm Z_1 = (0,0,1)^T\\
        & \bm Z_2 = (\sin i_M\sin\Omega_M,-\sin i_M\cos\Omega_M,\cos i_M)^T
    \end{aligned}   
\right.
\end{equation}
where $\varepsilon$ is the obliquity of the ecliptic plane with respect to the equatorial plane, $i_M$ and $\Omega_M$ are the inclination and the RAAN of the lunar orbit in GERS.
As $\sin i_M\approx 0.0894$ can be seen as small parameter, the symmetric matrix $\bm M$ can be expressed as follows:
\begin{equation}
    \bm M=\bm M_0+\bm M_1+\bm M_2
\end{equation}
where
\begin{equation}
    \bm M_0=
    \begin{pmatrix}
        0 & 0 & 0 \\
        0 & \omega_0\sin^2\varepsilon & 
            \omega_0\sin\varepsilon\cos\varepsilon \\
        0 & \omega_0\sin\varepsilon\cos\varepsilon &
            \omega_0\cos^2\varepsilon + \omega_1^*\\
    \end{pmatrix}
\end{equation}
\begin{equation}
    \bm M_1 = \frac{1}{2}\omega_2\sin2 i_M
    \begin{pmatrix}
        0 & 0 & \sin\Omega_M \\
        0 & 0 & -\cos\Omega_M \\
        \sin\Omega_M & 
        -\cos\Omega_M & 0\\
    \end{pmatrix}    
\end{equation}
\begin{equation}
    \bm M_2 = \frac{1}{2}\omega_2\sin^2i_M
    \begin{pmatrix}
        1-\cos2\Omega_M & -\sin2\Omega_M & 0 \\
        -\sin2\Omega_M & 1+\cos2\Omega_M & 0 \\
        0 & 0 & 0\\
    \end{pmatrix}
\end{equation}
with $\omega_1^*=\omega_1+\omega_2\cos^2i_M$.
The symmetric matrix $\bm M$ is composed of an autonomous part $\bm M_0$ and a time-dependent part $\bm M_1+\bm M_2$. The time-dependent part can be seen as perturbations with magnitudes dependent on $\sin i_M$.

\subsection{Intermediate solution}
The intermediate dynamical system is obtained by excluding the time-dependent tensor from Eq. (\ref{eq:motion}).
Similarly to \citet{Allan1964}, the intermediate solution is obtained through several steps listed below:
\begin{enumerate}[{\bf(1)}]
    \item \textbf{Eigenvalue and eigenvector analysis of $\bm M_0$}

    The Eigenvalues of $\bm M_0$ are $0=\lambda_1<\lambda_2<\lambda_3$
	\begin{equation}
		\lambda_{2,3}=\frac{1}{2}(\omega_0+\omega_1^*)\left(
		1\mp\sqrt{1-4\frac{\omega_0\omega_1^*}{(\omega_0+\omega_1^*)^2}\sin^2\varepsilon}
		\right)
	\end{equation}
	and the corresponding eigenvectors in GERS are
    \begin{equation}
    \left\{
    \begin{aligned}
        &\bm\eta_1 = C_1 \begin{pmatrix} 1 & 0 & 0 \end{pmatrix}^T \\
        &\bm\eta_2 = C_2 \begin{pmatrix} 
            0&\cos\theta&-\sin\theta\end{pmatrix}^T \\
        &\bm\eta_3 = C_3\begin{pmatrix} 
            0&\sin\theta&\cos\theta
        \end{pmatrix}^T
    \end{aligned}
    \right.
    \end{equation}
    where $C_j\neq 0$ for $j=1,2,3$ are arbitrary constants, and $\theta$ is a function of semi-major axis determined by
    \begin{equation}
        \tan\theta = \frac{\omega_0\sin\varepsilon\cos\varepsilon}
        {\lambda_3-\omega_0\sin^2\varepsilon}
    \end{equation}
	Fig. \ref{fig:lambda_theta} illustrates the eigenvalues $\lambda_{2,3}$ and angle $\theta$ as functions of semimajor axis for readers' convenience of reading off their magnitudes.
	\begin{figure}[htbp]
	\centering
		\includegraphics[width=0.45\textwidth]{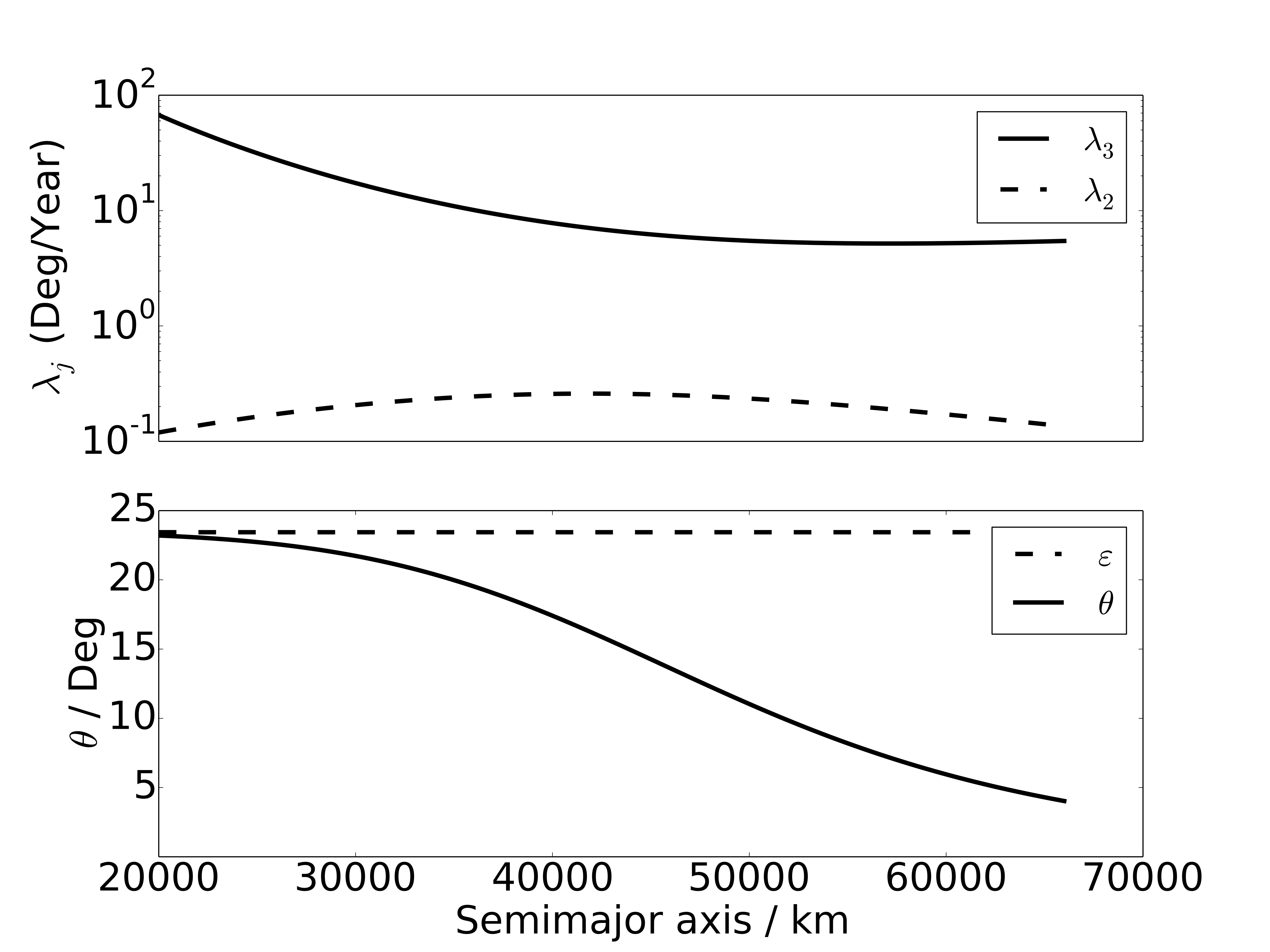}
		\caption{The eigenvalues $\lambda_{2,3}$ and angle $\theta$ as functions of semimajor axis.}
		\label{fig:lambda_theta}
	\end{figure}

    \item \textbf{Determination of the intermediate reference system}

    Let $C_{1,2,3} = 1$, then $\bm\eta_{1,2,3}$ form a right-hand Geocentric Intermediate Reference System (GIRS, $O\text{\,-\,}xyz$), in which $Ox$-axis points to the equinox, and $Oz$-axis points to the north with an angle $\theta$ from the ecliptic pole ($OZ_1$).
    Then the $O\text{\,-\,}xy$ plane is the local Laplace plane with an inclination $\alpha=\varepsilon-\theta$ to the equator.

    \item \textbf{Solve the equation of motion in the intermediate reference system analytically}

    In the $O\text{\,-\,}xyz$ reference system, the equation of motion is simplified to
    \begin{equation}
    \left\{
    \begin{aligned}
        & \dot x = (\lambda_3-\lambda_2)yz \\
        & \dot y = (\lambda_1-\lambda_3)xz \\
        & \dot z = (\lambda_2-\lambda_1)xy \\
    \end{aligned}
    \right.
    \end{equation}
    where $(x,y,z)^T$ is the coordinate of $\bm R$ in GIRS, and note that $\lambda_1=0$.
    There are two first integrals of the intermediate dynamical system:
    \begin{equation}
    \left\{
        \begin{aligned}
            & \bm R\cdot \bm R = x^2+y^2+z^2=1 \\
            & \bm R\cdot \bm M_0\bm R = \lambda_1x^2 + \lambda_2y^2 + \lambda_3z^2\triangleq\lambda \\
        \end{aligned}
    \right.
    \end{equation}
\end{enumerate} 

Finally, the solution is expressed analytically in two cases, depending on $\lambda$:
\begin{enumerate}[(a)]
    \item If $\lambda\in[0,\lambda_2]$, then 
    \begin{equation}
    \left\{
    \begin{aligned}
        &x = \sign(x)\sqrt{\frac{\lambda_3-\lambda}{\lambda_3}}\dn u \\
        &y = \sqrt{\frac{\lambda}{\lambda_2}}\sn u \\
        &z = \sqrt{\frac{\lambda}{\lambda_3}}\cn u\\
    \end{aligned}
    \right.
    \label{eq:solution_xyz_1}
    \end{equation}
    where $\dn$, $\sn$ and $\cn$ are Jacobi elliptic functions with $\kappa$ the elliptic modulus defined as follows: 
    \begin{equation}
        \kappa = \kappa_1 = 
        \sqrt{\frac{\lambda(\lambda_3-\lambda_2)}{\lambda_2(\lambda_3-\lambda)}}\in[0,1]
    \label{eq:kappa_1}
    \end{equation}
	and $u$ a linear function in time
	\begin{equation}
		u = u_0 - \sign(x)\sqrt{\lambda_2(\lambda_3 - \lambda)}t
	\end{equation}
	where $u_0$ is the initial value of $u$.
	The function $\sign(x)$ equals to 1 for positive $x$, $-1$ for negative $x$, and zero for $x=0$, and it is determined by the initial value and does not change along the solution trajectory.
    \item If $\lambda\in[\lambda_2,\lambda_3]$, then
    \begin{equation}
    \left\{
    \begin{aligned}
        & x = \sqrt{\frac{\lambda_3-\lambda}{\lambda_3}}
            \cn u \\
        & y = \sqrt{\frac{\lambda_3-\lambda}{\lambda_3-\lambda_2}}
            \sn u \\
        & z = \sign(z)\sqrt{\frac{\lambda}{\lambda_3}}\dn u
    \end{aligned}
    \right.
    \label{eq:solution_xyz_2}
    \end{equation}
    and the corresponding $(\kappa,u)$ are
    \begin{equation}
        \kappa = \kappa_2 =
        \sqrt{\frac{\lambda_2(\lambda_3-\lambda)}{\lambda(\lambda_3-\lambda_2)}}\in[0,1]
    \label{eq:kappa_2}
    \end{equation}
	\begin{equation}
		u = u_0 - \sign(z)\sqrt{\lambda(\lambda_3 - \lambda_2)}t
	\end{equation}
\end{enumerate} 

\subsection{Geometric view of the intermediate solution}
\begin{figure}[htbp]
    \centering
    \includegraphics[width=0.4\textwidth]{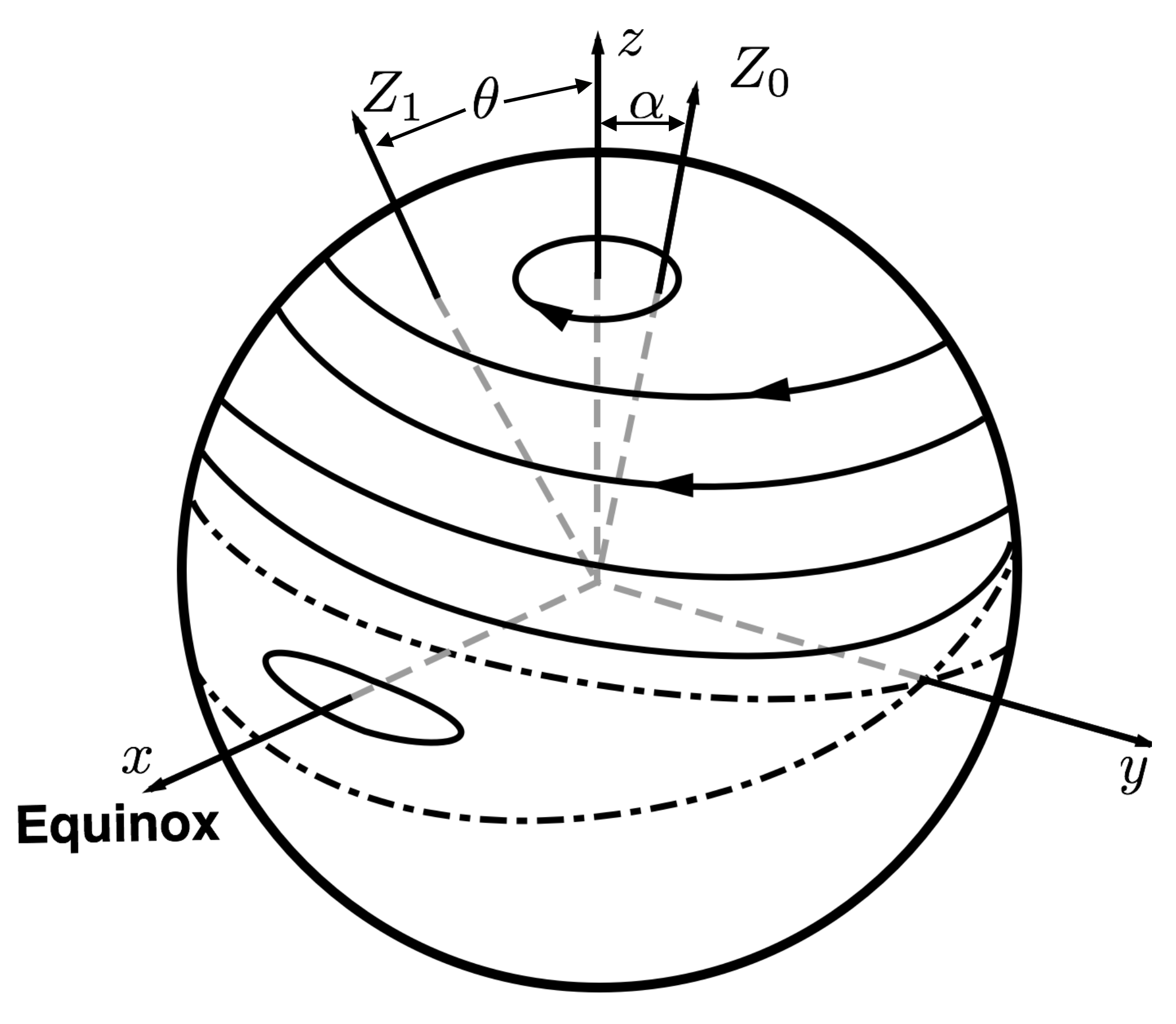}
    \caption{
	Geometric illustration of the intermediate solution \citep{Allan1964}. The $\bm Z_0$-axis alongs the Earth's rotating axis, the $\bm Z_1$-axis directs to the ecliptic pole. The $\bm z$-axis locates in the plane individuated by $\bm Z_0$ and $\bm Z_1$, with an angle $\theta$ to $\bm Z_1$ and an angle $\alpha$ to $\bm Z_0$.}
    \label{fig:intermediate_solution}
\end{figure}
As showed in Fig. \ref{fig:intermediate_solution}, the pole of the orbital plane precesses around the axis $Ox$ or $Oz$ for case (a) and (b) respectively, and they are separated by the separatrices (dash-dot line, corresponds to $\lambda = \lambda_2$) connected to the $Oy$-axis.
In addition, common orbits with small and medium inclination (with respect to the equator) are of case (b).

\subsection{Equation of perturbed motion}
\label{sec:equation_of_motion}
The equation of motion considering the time-dependent tensor can be expressed in GIRS as follows:
\begin{equation}
    \left\{
        \begin{aligned}
            \dot x \approx
                &  (\lambda_3-\lambda_2)yz+\epsilon\Big[
                    xy\sin\Omega_M\cos\theta \\
                & + xz\sin\Omega_M\sin\theta 
                  - y^2\cos\Omega_M\cos2\theta \\
                & - 2yz\cos\Omega_M\sin2\theta
                  + z^2\cos\Omega_M\cos2\theta\Big]\\
            \dot y \approx
                & -\lambda_3xz - \epsilon\Big[
                    x^2\sin\Omega_M\cos\theta \\
                & + yz\sin\Omega_M\sin\theta 
                  - xy\cos\Omega_M\cos2\theta \\
                & - z^2\sin\Omega_M\cos\theta 
                  - xz\cos\Omega_M\sin2\theta
            \Big]\\
            \dot z \approx
                & \lambda_2xy - \epsilon\Big[
                    x^2\sin\Omega_M\sin\theta \\
                & - y^2\sin\Omega_M\sin\theta 
                  - xy\cos\Omega_M\sin2\theta \\
                & + yz\sin\Omega_M\cos\theta 
                  + xz\cos\Omega_M\cos2\theta
            \Big]
        \end{aligned}
    \right.
\label{eq:perturbed_equation}
\end{equation}
where $\epsilon = \omega_2\sin i_M\cos i_M$, and the second order tensor $\bm M_2$ is ignored.

The phase space of system (\ref{eq:perturbed_equation}) is two dimensional considering that the first integral $\bm R\cdot\bm R = 1$ still holds.
Then the equation of the perturbed motion is established based on the theory of constant variation. 
Specifically, $\kappa$ is a first integral (or constant of motion) of the unperturbed dynamical system, and it determines the integral curve (i.e. solid circles with an arrow in Fig. \ref{fig:intermediate_solution}), while variable $u$ (linear function in time) determines the location on the curve at a specific time.
Taking into account of the perturbation, $\kappa$ turns to a slowly changing variable, and $u$ becomes quasi-linear with time.
According to the variable transformation $(x,y,z)\to(\lambda,u)\to(\kappa,u)$, the equation of the perturbed motion is derived easily, as outlined below.

Concentrating on case (b), the explicit expression of the transformation $(x,y)\to(\kappa,u)$ is
\begin{equation}
\left\{
\begin{aligned}
    & x = \Lambda(\kappa)\kappa\cn u \\
    & y = \sqrt{\frac{\lambda_3}{\lambda_3-\lambda_2}}
        \Lambda(\kappa)\kappa\sn u \\
\end{aligned}
\right.
\label{eq:transform_b}
\end{equation}
and the phase angle $w$ is defined by
\begin{equation}
	w=\frac{\pi u}{2K(\kappa)}
\end{equation}
where $K$ is the complete elliptic integral, and $\Lambda$ is the function of $\kappa$ ($\kappa\in(0,1)$) in the following form:
\begin{equation}
\begin{aligned}
    \Lambda(\kappa)
    &=\sqrt{\dfrac{\lambda_3-\lambda_2}{\lambda_3\kappa^2+\lambda_2(1-\kappa^2)}}\\
    &\in\left(\,\sqrt{\dfrac{\lambda_3-\lambda_2}{\lambda_3}}, \sqrt{\dfrac{\lambda_3-\lambda_2}{\lambda_2}}\,\right)
\end{aligned}
\end{equation}
The equation after transformation (\ref{eq:transform_b}) is defined by
\begin{equation}
\begin{aligned}
    \begin{pmatrix}
        \dot\kappa \\ \\ \dot u
    \end{pmatrix}
    &= \left(\frac{\partial(x,y)}{\partial(\kappa, u)}\right)^{-1}
    \begin{pmatrix}
        \dot x \\ \\ \dot y
    \end{pmatrix}\\
    &= \dfrac{1}{\det\left(\frac{\partial(x,y)}{\partial(\kappa, u)}\right)}
    \begin{pmatrix}
        \dfrac{\partial y}{\partial u} &
        -\dfrac{\partial x}{\partial u} \\
        & \\
        -\dfrac{\partial y}{\partial\kappa} &
        \dfrac{\partial x}{\partial\kappa}\\
    \end{pmatrix}
    \begin{pmatrix}
        \dot x \\ \\ \dot y
    \end{pmatrix}
\end{aligned}
\label{eq:def_variational_equation_b}
\end{equation}
where 
\begin{equation}
	\det\left(\frac{\partial(x,y)}{\partial(\kappa,u)}\right) = 
	\sqrt{\frac{\lambda_3}{\lambda_3-\lambda_2}}
	\Big[1+\Lambda^2(\kappa)\kappa^2\Big]\Lambda^2(\kappa)\kappa\dn u
\end{equation}
is the Jacobi determinant, and the factor $\kappa$ leads to vanishing divisor at $\kappa=0$ for the expression of $\dot u$ (see below).

After a series of cumbersome derivations and introducing the Fourier series of elliptic functions, Eq. (\ref{eq:def_variational_equation_b}) reduces to Eqs. (\ref{eq:variational_equation_b_kappa}) and (\ref{eq:variational_equation_b_u}):
\begin{equation}
    \begin{aligned}
        \dot\kappa = & \epsilon \bigg[
		A_{1,-1}(\kappa)\cos(w-\Omega_M) 
        + A_{1,1}(\kappa)\cos(w+\Omega_M) \\
        & + A_{0,1}(\kappa)\sin\Omega_M 
        + A_{2,-1}(\kappa)\sin(2w-\Omega_M) \\
        & + A_{2,1}(\kappa)\sin(2w+\Omega_M) + \mathcal O(\kappa^2)\bigg]
		+ \mathcal O(\epsilon^2)
    \end{aligned}
\label{eq:variational_equation_b_kappa}
\end{equation}
and
\begin{equation}
	\dot u = -\sign(z)\sqrt{\lambda_2\lambda_3}\Lambda(\kappa) + \mathcal O(\epsilon/\kappa)
	\label{eq:variational_equation_b_u}
\end{equation}
For convenience, we replace $u$ with the phase angle $w$ as an independent variable, then
\begin{equation}
	\begin{aligned}
	    \dot w =& \frac{\pi}{2K}\dot u - \frac{\pi u}{2K^2}\frac{\ud K}{\ud\kappa}\dot\kappa\\
		= & -\sign(z)\sqrt{\lambda_2\lambda_3}\frac{\Lambda\pi}{2K} + 
	    \mathcal O(\epsilon/\kappa)
	\end{aligned}
\label{eq:variational_equation_b_w}
\end{equation}
Note that the neglected terms $\mathcal O(\kappa^2)$ in Eq. (\ref{eq:variational_equation_b_kappa}) are not all the second order terms of $\kappa$, but only those from the Fourier series of Jacobi's elliptic functions, which are of smaller amplitudes and higher frequencies than the leading terms. 
In addition, the changing rate of angle $u$ or $w$ is dominated by the zeroth order term of $\epsilon$ (zeroth order secular term) and the neglected terms in $\mathcal O(\epsilon/\kappa)$ are first order periodic terms (with the period of 18.6 years).
Due to the singularity at $\kappa=0$, the above simplifications are valid for $\kappa$ that is not so small according to our numerical tests.
And analytic solution for extremely small $\kappa$ is developed in another form in order to avoid the small divisor problem.

By introducing the truncated power series of elliptic integral $K$, the amplitudes $A_{i,j}$'s in Eq. (\ref{eq:variational_equation_b_kappa}) can be approximated expressed as follows:
\begin{equation}
\left\{
    \begin{aligned}
        A_{1,\pm 1} \approx 
		& \frac{\cos2\theta\mp\cos\theta}{2\Lambda}\\
        A_{0,1} \approx 
		& \frac{\kappa\sign(z)}{16\sqrt{1-\kappa^2\Lambda^2}}\\
		& \times \left[8-3\kappa^2-(8-\kappa^2)\sqrt{\frac{\lambda_3}{\lambda_3-\lambda_2}}
		\right]\sin\theta \\
        A_{2,\pm 1} \approx &
        \frac{\kappa\sign(z)}{4\sqrt{1-\kappa^2\Lambda^2}}
        \Bigg[\pm
            \left(1+\sqrt{\frac{\lambda_3}{\lambda_3-\lambda_2}}
            \right)\sin\theta \\
         & + \left(1-2\sqrt{\frac{\lambda_3}{\lambda_3-\lambda_2}}
            \right)\sin2\theta
        \Bigg]
    \end{aligned}
\right.
\end{equation}
We do not further simplify the amplitudes by expanding $\Lambda$ and $\sqrt{1-\kappa^2\Lambda^2}$, because the ratio of eigenvalues, $\frac{\lambda_2}{\lambda_3}$, is small, to the extent that the expansion would cause severe loss of accuracy.

\subsection{Model validation}
\begin{figure}[htbp]
    \centering
    \includegraphics[width=0.48\textwidth]{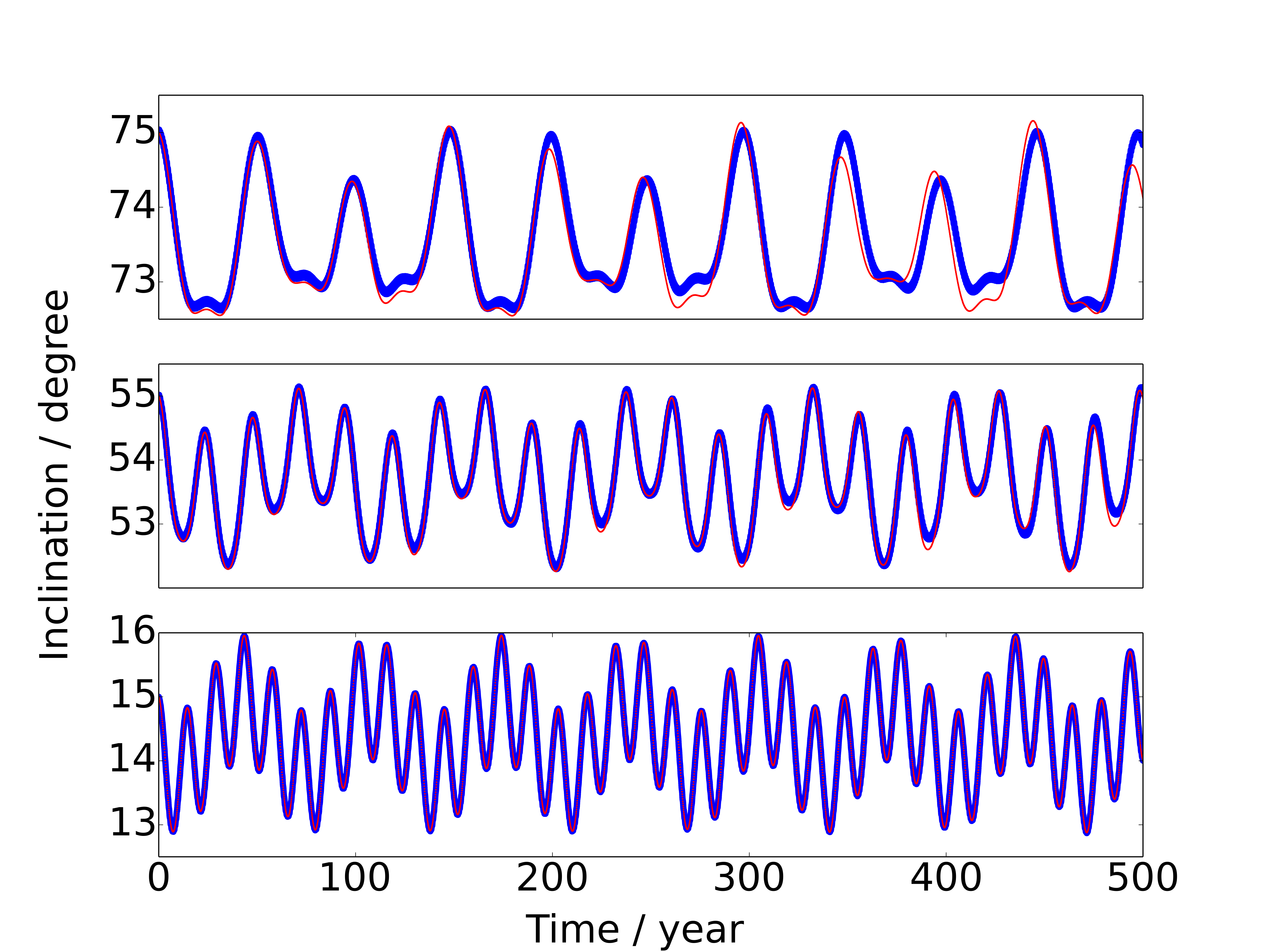}
    \caption{Time series of inclination for ($a={26554}{\km}$). Thicker blue: generated by STELA; thinner red: generated by the doubly averaged model. The initial epoch is J2000.0, the initial Keplerian elements are $a={26554}{\km}$, $e=0$, $i={15}{\degree}$ (top), ${55}{\degree}$ (middle), ${75}{\degree}$ (bottom) and $\Omega={0}{\degree}$}
    \label{fig:model_verification_26554}
\end{figure}

The validation of the doubly averaged model in studying the long-term evolution of distant circular orbits has been confirmed in a number of papers \citep{Allan1964,Rosengren2014a,Ulivieri2013}.
We verify the model by comparing the time series of inclination due to Eq. (\ref{eq:motion}) and the results from the STELA software \citep{stela}.
The Earth's zonal harmonics up to 4th degree ($J_2$, $J_3$ and $J_4$ terms) and the luni-solar gravity (based on the simplified Meeus and Brown theory, a highly accurate analytic model) are taken into consideration in the STELA simulation.
Fig. \ref{fig:model_verification_26554} illustrates three of our verification examples, the initial (at epoch J2000.0) mean Keplerian elements are $a={26554}{\km}$, $e=0$, $i={15}{\degree}, {55}{\degree}, {75}{\degree}$, and $\Omega={0}{\degree}$.

%% file: Solution.tex
\section{Analytic approach}
\label{sec:solution}

\subsection{General solution for the perturbed motion}
Observing the Eqs. (\ref{eq:variational_equation_b_kappa}) and (\ref{eq:variational_equation_b_w}), it is obvious that $\kappa$ varies quasi-periodically with the amplitude of the order of $\mathcal O(\epsilon)$.
Then it is reasonable to expand the right-hand side of both the equations in the vicinity of some mean value of $\kappa$ (denoted by $\bar\kappa$ and to be determined later) and retaining the leading terms only:
\begin{equation}
    \begin{aligned}
        \dot\kappa = & \epsilon \bigg[
		A_{1,-1}(\bar\kappa)\cos(w-\Omega_M) 
        + A_{1,1}(\bar\kappa)\cos(w+\Omega_M) \\
        & + A_{0,1}(\bar\kappa)\sin\Omega_M 
        + A_{2,-1}(\bar\kappa)\sin(2w-\Omega_M) \\
        & + A_{2,1}(\bar\kappa)\sin(2w+\Omega_M) \bigg]
		+ \mathcal O(\epsilon^2)
    \end{aligned}
\end{equation}
and
\begin{equation}
    \dot w = -\sign(z)\sqrt{\lambda_2\lambda_3}
    \Lambda(\bar\kappa)\left(1-\frac{1}{4}\bar\kappa^2\right) + 
    \mathcal O(\epsilon/\kappa) 
\end{equation}

Denote the variables' initial values by $w_0$ and $\kappa_0$, the approximate solutions are obtained as
\begin{equation}
    w = w_0 -\sign(z)\sqrt{\lambda_2\lambda_3}
    \Lambda(\bar\kappa)\left(1-\frac{1}{4}\bar\kappa^2\right)(t-t_0) \\
\label{eq:w_analytic}
\end{equation}
and
\begin{equation}
	\begin{aligned}
		\kappa =& \bar\kappa + \epsilon\bigg[
			\bar B_{1,-1}\sin(w-\Omega_M) + \bar B_{1,1}\sin(w+\Omega_M) \\
	        & - \bar B_{0,1}\cos\Omega_M - \bar B_{2,-1}\cos(2w-\Omega_M) \\
	        & - \bar B_{2,1}\cos(2w+\Omega_M) \bigg]
	\end{aligned}
\label{eq:kappa_analytic}
\end{equation}
where $\bar B_{i,j}=B_{i,j}(\bar\kappa)=A_{i,j}(\bar\kappa) / (i\dot w(\bar\kappa) + j\dot\Omega_M)$ are constants.
The mean $\kappa$ is obtained by removing all the periodic terms from the osculating value at $t=t_0$, i.e.
\begin{equation}
	\begin{aligned}
		\bar\kappa =& \kappa_0 - \epsilon\bigg[
			\bar B_{1,-1}\sin(w_0-\Omega_M^0) 
	        + \bar B_{1,1}\sin(w_0+\Omega_M^0) \\
	        & - \bar B_{0,1}\cos\Omega_M^0 
	        - \bar B_{2,-1}\cos(2w_0-\Omega_M^0) \\
	        & - \bar B_{2,1}\cos(2w_0+\Omega_M^0) \bigg]
	\end{aligned}
\label{eq:kappa_star}
\end{equation}
with $\Omega_M^0=\Omega_M(t_0)$ the initial value. 
In Eq. (\ref{eq:kappa_star}), $\bar B_{i,j}$'s can be replaced by $B_{i,j}(\kappa_0)$ to the first order.

Finally, the analytic model for the perturbed motion in terms of variables $(\kappa,w)$ is expressed by their approximate solutions (Eqs. (\ref{eq:w_analytic}) and (\ref{eq:kappa_analytic})) along with Eq. (\ref{eq:kappa_star}).

\begin{remark}
    The assumption on the variation range of $\kappa$ does not hold when resonance happens, i.e. when $i\dot w + j\dot\Omega_M \approx 0$. As a result, the approximate general analytic solution fails for resonant cases.
\label{rmk:resonance}
\end{remark}

\subsection{Evolution near the classical Laplace plane}
As presented in section \ref{sec:equation_of_motion}, the equation of the perturbed motion may not be appropriate for the orbits in the neighborhood of classical Laplace plane, i.e. when $\kappa\approx 0$.
Traditionally, the circular orbits on the Laplace plane are frozen (i.e. the orbital plane has no secular change).
As a result, here we call the orbits near the classical Laplace plane the \textit{quasi-frozen orbits}.

The analytic approach on the quasi-frozen orbits stands on linearizing Eq. (\ref{eq:perturbed_equation}) at $\kappa=0$.
Note again that in this paper, we consider case (b) only.
Then it reduces to
\begin{equation}
\left\{
    \begin{aligned}
        \dot x =& (\lambda_3-\lambda_2)y +
            \epsilon\cos2\theta\cos\Omega_M +
            \mathcal O(\epsilon\kappa,\kappa^2)\\
        \dot y =& -\lambda_3x +
            \epsilon\cos\theta\sin\Omega_M +
            \mathcal O(\epsilon\kappa,\kappa^2)\\
    \end{aligned}
\right.
\end{equation}
Let $p=\sin\Omega_M$ and $q=\cos\Omega_M$, neglect higher-order terms:
\begin{equation}
    \frac{\ud}{\ud t}
    \begin{pmatrix}
        x \\ y \\ p \\ q
    \end{pmatrix}
    =
    \begin{pmatrix}
        0 & \lambda_3-\lambda_2 & 0 & \epsilon\cos2\theta \\
        -\lambda_3 & 0 & \epsilon\cos\theta & 0 \\
        0 & 0 & 0 & \dot\Omega_M \\
        0 & 0 & -\dot\Omega_M & 0 
    \end{pmatrix}
    \begin{pmatrix}
        x \\ y \\ p \\ q
    \end{pmatrix}
\end{equation}
The eigenvalues of the coefficient matrix are $\big\{\pm\ii\dot\Omega_M, \pm\ii\nu_2\big\}$, where $\nu_2 = \sqrt{\lambda_3(\lambda_3-\lambda_2)}$.
If $\dot\Omega_M^2\neq\nu_2^2$, the solution takes the following form:
\begin{equation}
\left\{
    \begin{aligned}
        & x = C_1\cos\Omega_M + S_1\sin\Omega_M +
            C_2\cos\nu_2t + S_2\sin\nu_2t \\
        & y = D_1\cos\Omega_M + E_1\sin\Omega_M + 
            D_2\cos\nu_2t + E_2\sin\nu_2t \\
    \end{aligned}
\right.
\label{eq:quasi_frozen_solution_2}
\end{equation}
The coefficients in Eq. (\ref{eq:quasi_frozen_solution_2}) satisfy
\begin{equation}
\left\{
    \begin{aligned}
        & -\dot\Omega_MC_1 = (\lambda_3-\lambda_2)E_1 \\
        & \dot\Omega_ME_1 = -\lambda_3C_1 \\
        & \dot\Omega_MS_1 = (\lambda_3-\lambda_2)D_1 + 
            \epsilon\cos2\theta\\
        & \dot\Omega_MD_1 = \lambda_3S_1 - 
            \epsilon\cos\theta\\
        & -\nu_2C_2=(\lambda_3-\lambda_2)E_2 \\
        & -\nu_2D_2=-\lambda_3S_2 \\
        & x^0 = C_1\cos\Omega_M^0 + S_1\sin\Omega_M^0 + C_2\\
        & y^0 = D_1\cos\Omega_M^0 + E_1\sin\Omega_M^0 + D_2\\
    \end{aligned}
\right.
\end{equation}
where $(x^0, y^0, z^0)$ and $\Omega_M^0$ stand for the initial values.
Then
\begin{equation}
\left\{
    \begin{aligned}
        & C_1 = E_1 = 0 \\
        & S_1 = \frac{\epsilon}{\dot\Omega_M^2-\nu_2^2}
            \Big[\dot\Omega_M\cos2\theta-
                (\lambda_3-\lambda_2)\cos\theta
            \Big] \\
        & D_1 = -\frac{\epsilon}{\dot\Omega_M^2-\nu_2^2}
            \Big[\dot\Omega_M\cos2\theta-
                \lambda_3\cos\theta
            \Big] \\
        & C_2 = x^0 - S_1\sin\Omega_M^0,\quad
            D_2 = y^0 - D_1\cos\Omega_M^0 \\
        & E_2 = -\sqrt{\frac{\lambda_3}{\lambda_3-\lambda_2}}C_2,
            \quad 
            S_2 = \sqrt{\frac{\lambda_3-\lambda_2}{\lambda_3}}D_2
    \end{aligned}
\right.
\label{eq:quasi_frozen_coeff_2}
\end{equation}
Finally, the orbital plane evolution of the quasi-frozen orbits for case (b) is approximately described by the analytic expression of Eqs. (\ref{eq:quasi_frozen_solution_2}), (\ref{eq:quasi_frozen_coeff_2}), and 
\begin{equation}
	z=\sign(z^0)\sqrt{1-x^2-y^2}
\end{equation}

\section{Numerical tests on the analytic model}
\label{sec:numerical}
The analytic expression is verified by comparing with numerical solution of the doubly averaged model, i.e. Eq. (\ref{eq:motion}).

Both the numerical and analytic solutions are converted back to Keplerian elements for the clarity of physical meaning.
Then the time series of deviation in the inclination and RAAN (noted by $\Delta i (t)$ and $\Delta \Omega (t)$ respectively) of the analytic results from the numerical ones is obtained.
At last, we calculate the root mean square (RMS) of $\Delta X$ by
\begin{equation}
    \big\Vert\Delta X(t)\big\Vert_2 = 
    \sqrt{\frac{1}{N+1}\sum_{k=0}^{N} \big\vert\Delta X(t_k)\big\vert^2}
\end{equation}
with $X$ either the inclination or RAAN, $N=5000$ in our simulation, and $t_k = k\times 0.1$ year.

For a given semi-major axis, the comparison sample set is generated as follows:
\begin{itemize}
    \item Initial inclination: $i\in({0}{\degree},{75}{\degree})$ with the step size being ${1}{\degree}$
    \item Initial RAAN: $\Omega\in({0}{\degree},{360}{\degree})$ with the step size being ${10}{\degree}$
    \item The initial epoch is fixed at J2000.0, so $\Omega_M$ is not sampled to cover $({0}{\degree},{360}{\degree})$. 
\end{itemize}
Then the Keplerian elements for the circular orbits of each sample are converted to vector expressions in GCRS (Geocentric Celestial Reference System), GERS, and GIRS.

The numerical integration of Eq. (\ref{eq:motion}) is applied in GERS, using the classical Runge-Kutta-Dormand-Prince 8(9) method. 
The time range of the integration is 500 years.

\begin{figure}[htbp]
    \centering
    \includegraphics[width=0.48\textwidth]{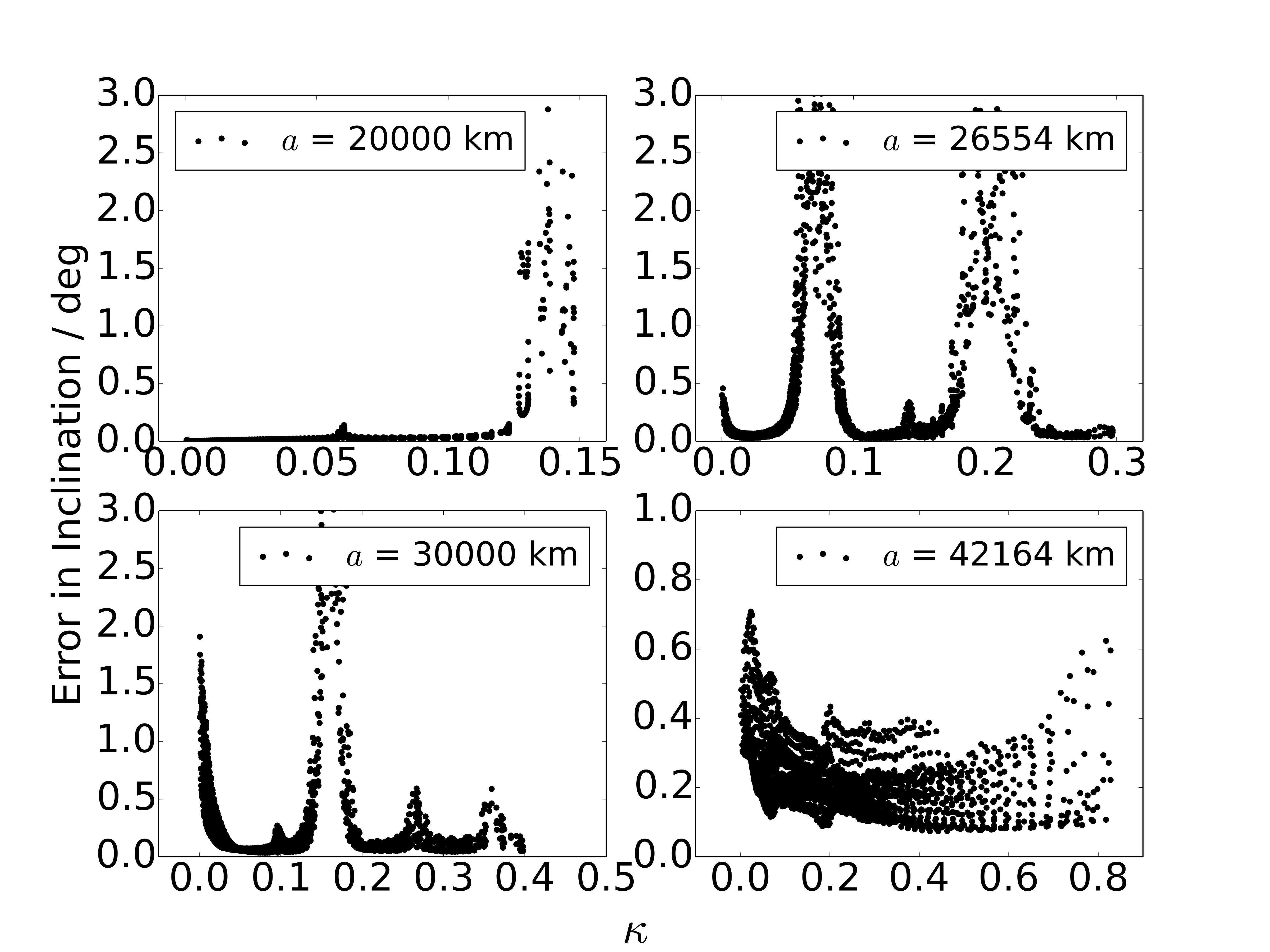}
    \caption{Error in inclination. The x-axis stands for initial value of $\kappa$ and the y-axis stands for $\big\Vert\Delta i(t)\big\Vert_2$.}
    \label{fig:inclination_error}
\end{figure}

\begin{figure}[htbp]
    \centering
    \includegraphics[width=0.48\textwidth]{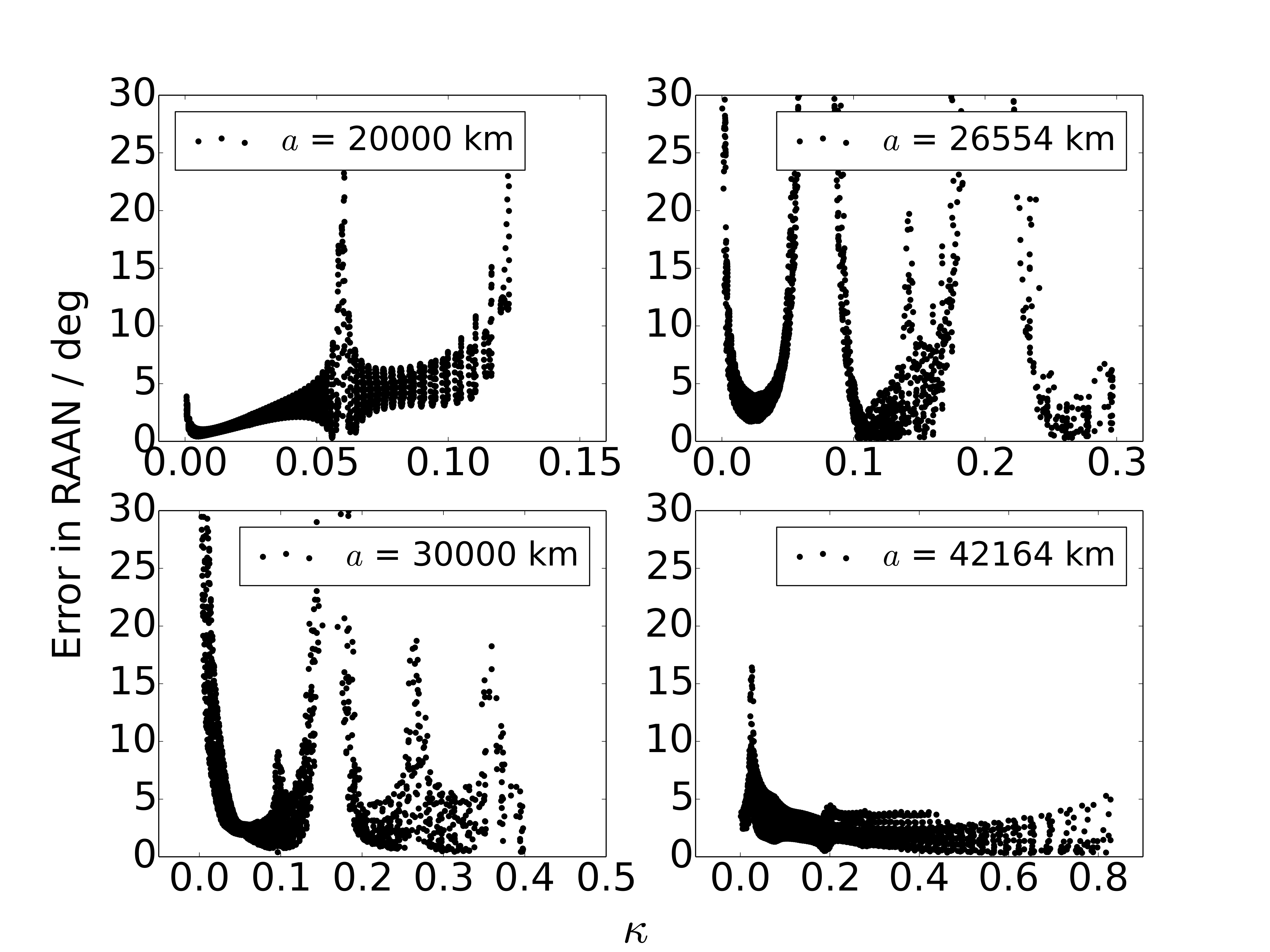}
    \caption{Error in RAAN. The x-axis stands for initial value of $\kappa$ and the y-axis stands for $\big\Vert\Delta \Omega(t)\big\Vert_2$.}
    \label{fig:raan_error}
\end{figure}

\begin{figure}[htbp]
    \centering
   \includegraphics[width=0.48\textwidth]{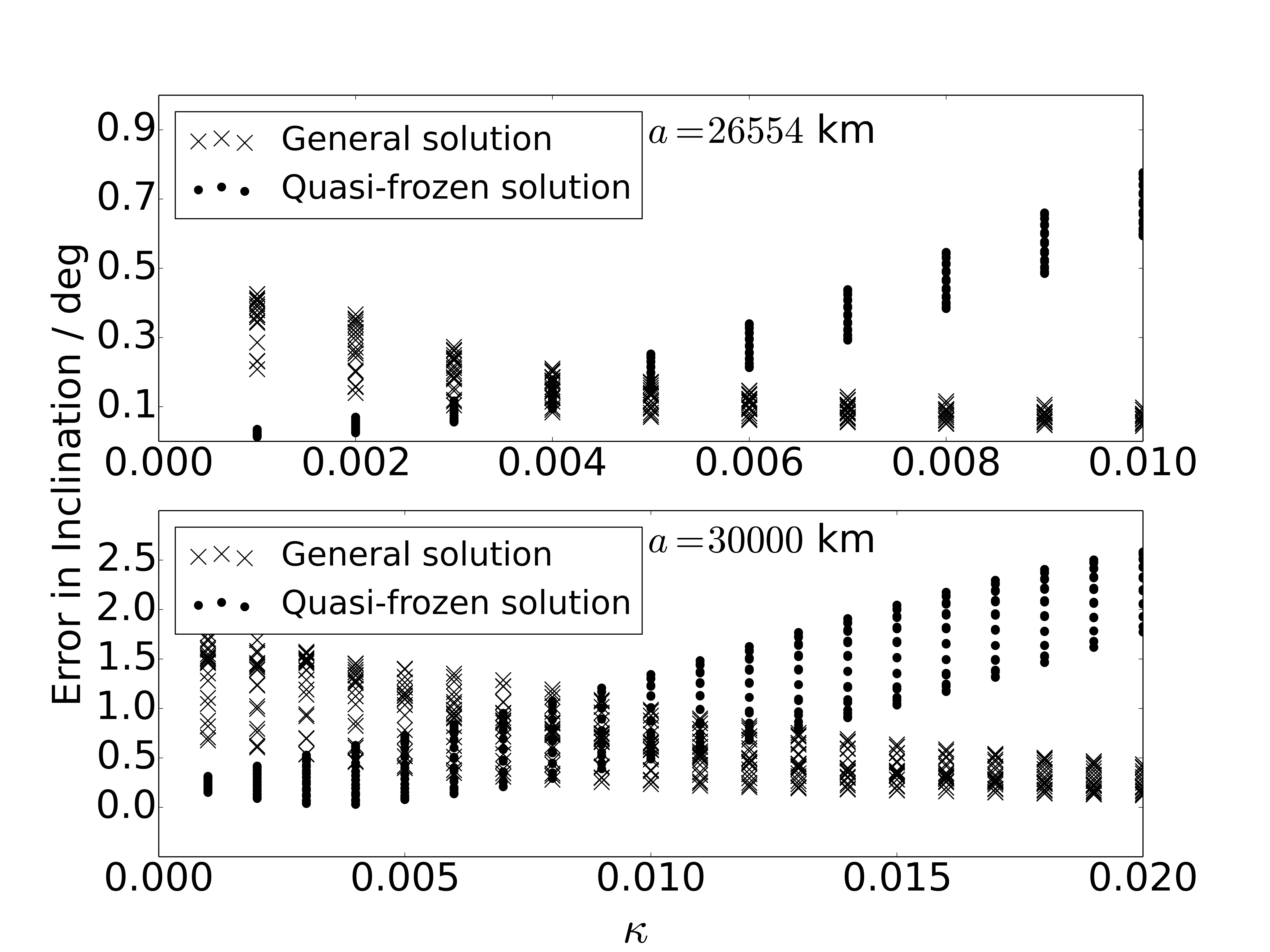}
    \caption{Error in inclination for small $\kappa$. The x-axis stands for initial value of $\kappa$ and the y-axis stands for $\big\Vert\Delta i(t)\big\Vert_2$.}
    \label{fig:inclination_error_quasi_frozen}
\end{figure}

\begin{figure}[htbp]
    \centering
    \includegraphics[width=0.48\textwidth]{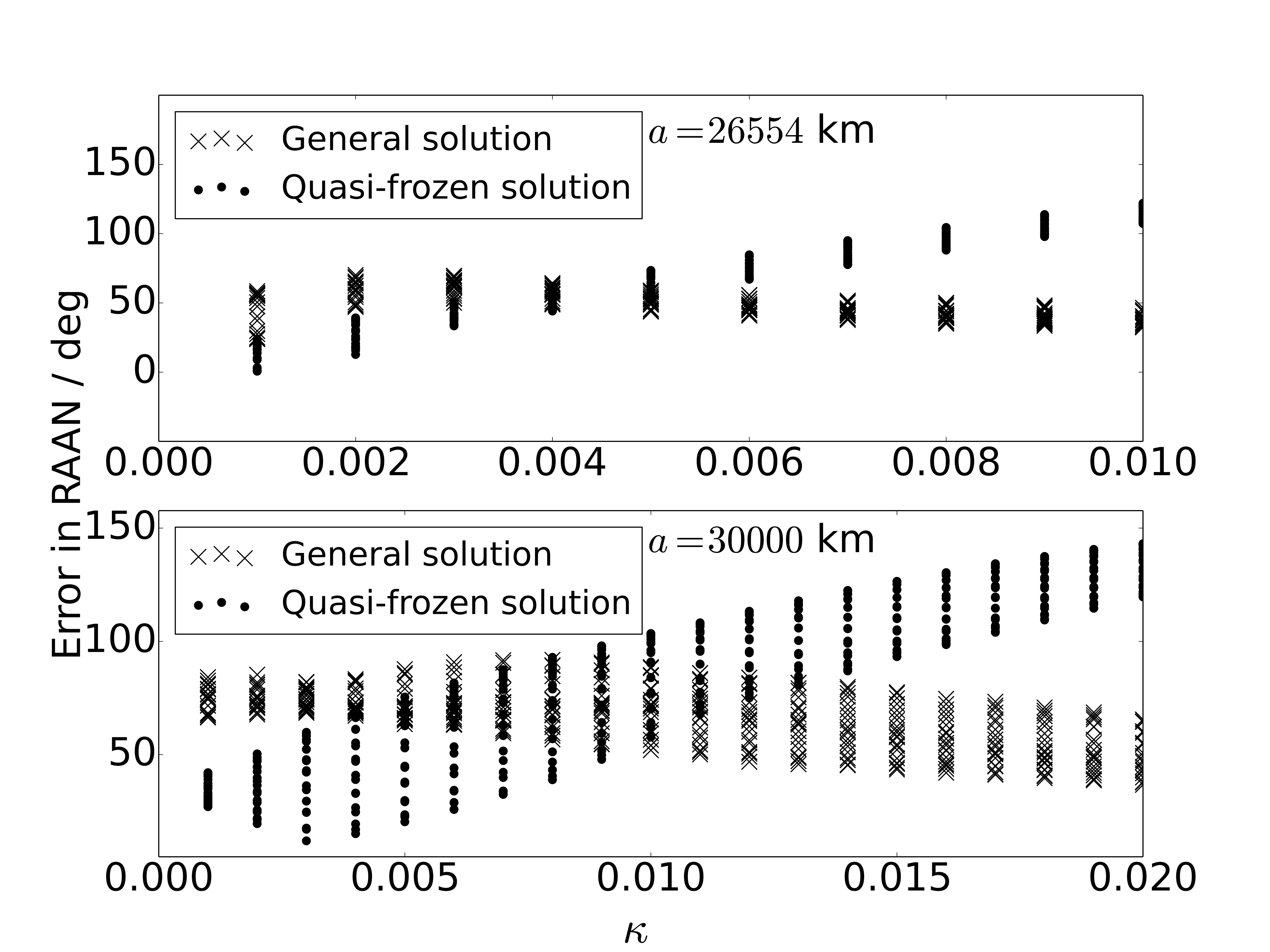}
    \caption{Error in RAAN for small $\kappa$. The x-axis stands for initial value of $\kappa$ and the y-axis stands for $\big\Vert\Delta \Omega(t)\big\Vert_2$.}
    \label{fig:raan_error_quasi_frozen}
\end{figure}

A two dimensional scatter is obtained for each semi-major axis (see Figs. \ref{fig:inclination_error} and \ref{fig:raan_error}).
The general analytic solution have comparatively large error when $\kappa\approx 0$.
This is rather obvious for the cases $a=26554\km$ and $a=30000\km$.
For small $\kappa$, we introduce the quasi-frozen analytic expression, i.e. Eqs. (\ref{eq:quasi_frozen_solution_2}) and (\ref{eq:quasi_frozen_coeff_2}).
The mean errors in inclination and RAAN for both analytic solutions are illustrated in Figs. \ref{fig:inclination_error_quasi_frozen} and \ref{fig:raan_error_quasi_frozen}, which show a better accuracy of quasi-frozen solution compared to the general analytic approach for small $\kappa$.

To note that the y-axis in Figs. \ref{fig:inclination_error} and \ref{fig:raan_error} are limited to no larger than ${3}{\degree}$ and ${30}{\degree}$ respectively, then some points are removed from the first three plots (specifically, points at $\kappa\approx 0.14$ for $a=20000{\km}$ , at $\kappa\approx 0.07$ and $\kappa\approx 0.2$ for $a=26554{\km}$ , at $\kappa\approx 0.15$ for $a=30000{\km}$), which corresponds to the resonances as mentioned in Remark \ref{rmk:resonance}.

An extended simulation is applied to cover the principle medium and high Earth orbits.
The samples are generated in the three dimensional initial value space: 
\begin{itemize}
     \item $a\in({20000}{\km},{45000}{\km})$ with the step size of ${500}{\km}$
     \item $i\in({0}{\degree},{75}{\degree})$ with the step size of ${1}{\degree}$
     \item $\Omega\in({0}{\degree},{360}{\degree})$ with the step size of ${10}{\degree}$
     \item The initial epoch is fixed at J2000.0
\end{itemize}
\begin{figure}[htbp]
    \centering
    \includegraphics[width=0.45\textwidth]{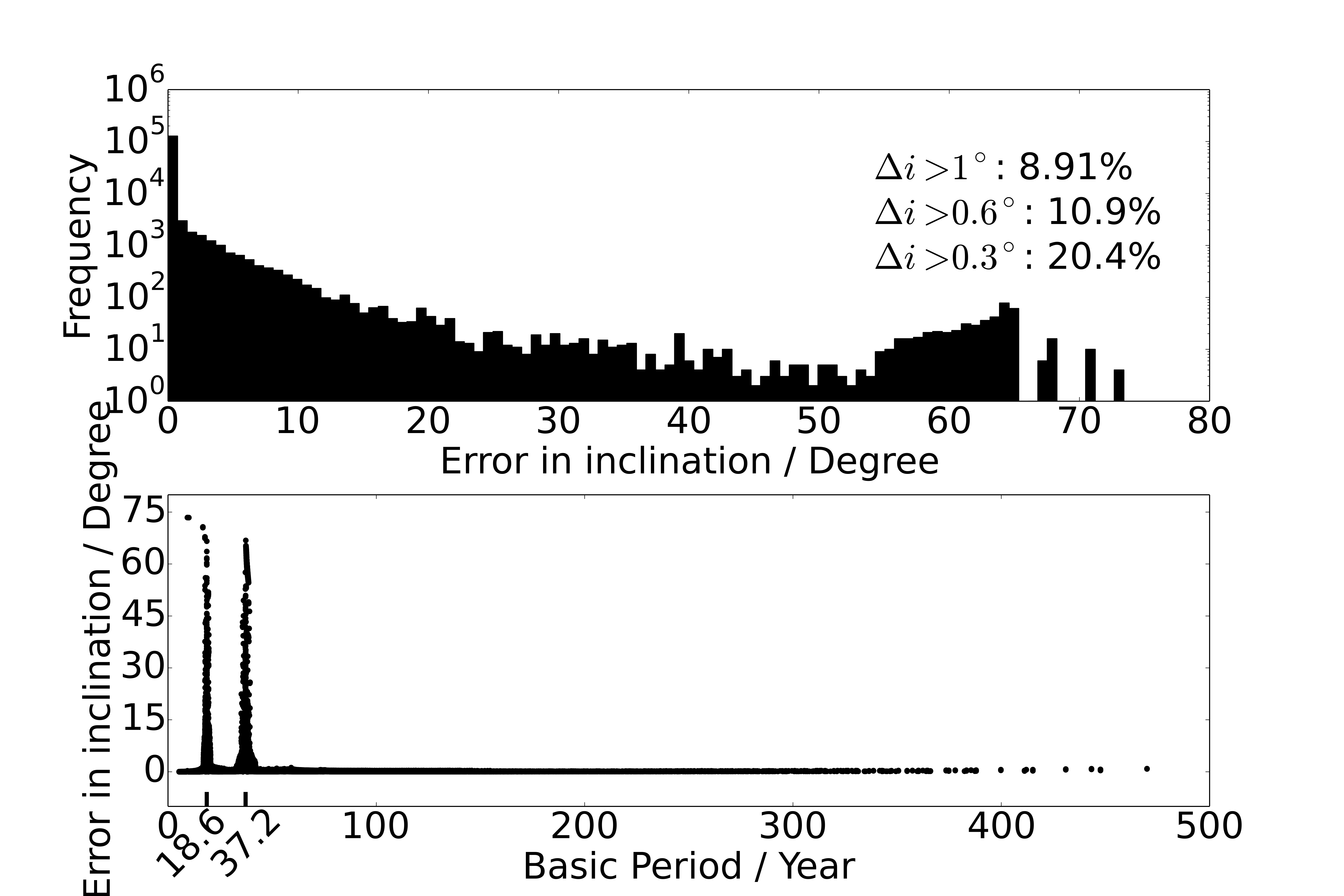}
    \caption{Histogram of error in inclination (above) and its relation with the basic period (bottom)}
    \label{fig:inclination_error_full}
\end{figure}
\begin{figure}[htbp]
    \centering
    \includegraphics[width=0.45\textwidth]{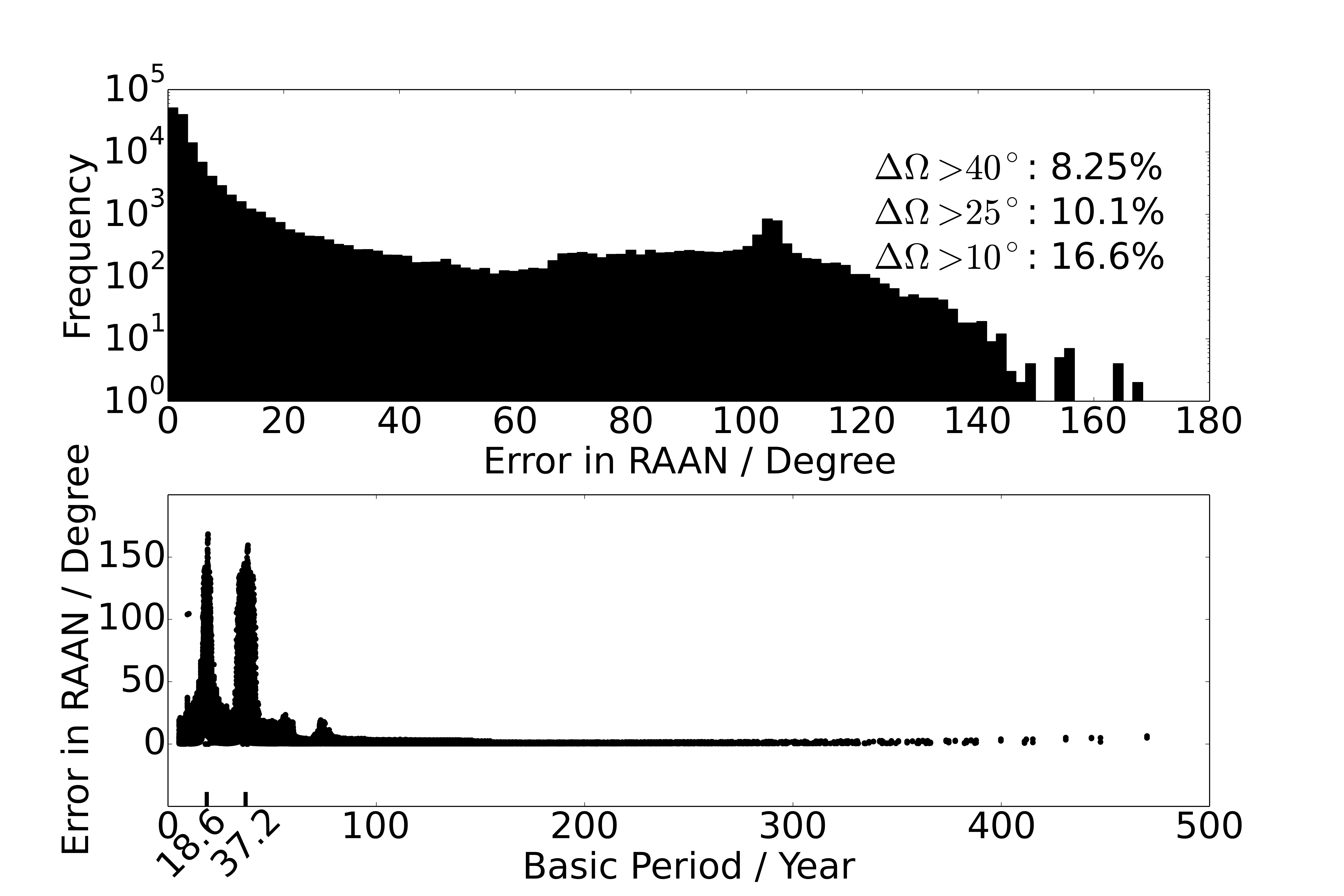}
    \caption{Histogram of error in RAAN (above) and its relation with the basic period (bottom)}
    \label{fig:raan_error_full}
\end{figure} 
Fig. \ref{fig:inclination_error_full} illustrates the histogram of errors in inclination (above), and its relation with the basic period by scatter plot (bottom). The errors in RAAN are showed in Fig. \ref{fig:raan_error_full}.
The histograms show that the percentage of samples with $\big\Vert\Delta i(t)\big\Vert_2 < {0.6}{\degree}$ and $\big\Vert\Delta\Omega\Vert_2 < {25}{\degree}$ is about ${90}{\%}$ among our sample set.
The bottom plots reveal a strong correlation between the failing analytic model and the resonances at $T_w\approx nT_M$ for $n=1,2$,
where $T_w = 2\pi / \dot w$ (depends on $a$ and $\kappa$) is the basic period, and $T_M \approx {18.6}$ year is the regression period of the lunar node.

%% file: Conclusion.tex
\section{Conclusion and discussion}
We have established an approximate analytic theory of the orbital plane evolution of circular Earth satellite orbits with high altitude.
The force model consists of the Earth's oblateness perturbation, and the luni-solar gravitational perturbations, with lunar orbital inclination to the ecliptic plane and the node regression taken into consideration.
A doubly averaged model is established using the Milankovitch elements. 
The approximate analytic expression for the long period orbital plane evolution is derived using the method of the variation of constants, on the basis of the classical analytic solution, in which the lunar node regression was omitted.
Appropriate amendments have been applied to the orbits close to the classical Laplace plane.
According to our analysis and verification, the analytic approach is valid for most medium and high Earth orbits with small eccentricity, except for the resonant cases.